\documentstyle[12pt]{article}
\vspace{-3cm}

\begin{document}

\title{\large\bfseries
PHYSICAL MODEL OF DIRAC ELECTRON. CALCULATION \\
OF ITS MASS AT REST AND OWN ELECTRIC AND MAGNETIC
INTENSITIES ON ITS MOMENT LOCATION.}
\vspace{-2.6cm}
\author{\large\slshape Josiph Mladenov Rangelov\, \\
Institute of Solid State Physics\,,\, Bulgarian Academy of Sciences\, \\
N\,72\, Tsarigradsko chaussee\,,\,1784 Sofia\,,\, Bulgaria }
\date{}
\maketitle

\vspace{-1.2cm}
\begin{abstract}
 The physical model (PhsMdl) of the relativistic quantized Dirac's electron
(DrEl) is proposed, in which one be regarded as a point-like (PntLk)
elementary electric charge (ElmElcChrg), taking simultaneously part in the
following four disconnected different motions : a/ in Einstein's relativistic
random trembling harmonic shudders as a result of momentum recoils (impulse
kicks), forcing the DrEl's PntLk ElmElcChrg at its continuous emission and
absorption of high energy (Hgh-Enr) stochastic virtual photons (StchVrtPhtns)
by its PntLk ElmElcChrg; b/ in Schrodinger's fermion harmonic oscillations of
DrEl's fine spread (FnSpr) ElmElcChrg, who minimizes the self-energy at a
rest of is an electromagnetic self-action between its continuously moving
FnSpr ElmElcChrg and proper magnetic dipole moment (MgnDplMmn) and the
corresponding potential and vector-potential; c/in nonrelativistic Furthian
quantized stochastic boson circular harmonic oscillations as a result of the
permanent electris or magnetic interaction of its well spread (WllSpr)
ElmElcChrg or proper MgnDplMmn with the electric intensity (ElcInt) and the
magnetic intensity (MgnInt) of the resultant quantized electromagnetic field
(QntElcMgnFld) of all the stochastic generated virtual photons (StchVrtPhtns)
within the fluctuating vacuum (FlcVcm); d/in Nweton's classical motion along
a clear-cut smooth thin line as a result of some interaction of its over
spread (OvrSpr) ElmElcChrg, MgnDplMmn or bare mass with the intensities of
some classical fields. If all the relativistic dynamical properties of the
DrEl are results of the participate of its FnSpr ElmElcChrg in the strong
correlated self-consistent Schrodinger's relativistic fermion vortical
harmonic oscillations, then all the quantized dualistic dynamical properties
of the SchEl are results of the participate of its WllSpr ElmElcChrg in the
nonrelativistic Furthian quantized stochastic boson circular harmonic
oscillations.

\end{abstract}

\section{Introduction.}

 The successful scientific research of some natural phenomenon is very often
connected with some necessary idealization and soma minimum simplification
of the phenomenon under investigation. Many marvellous phenomena and
remarkable properties of the sunstance have been described by help of the
powerful logic of the Quantum Theory (QngThr). The physical model (PhsMdl) of
some physical phenomenon presents as an actual ingradient of the physical
theory (PhsThr).  This is a scientific way for construction of the (PhsMdl)
of the Dirac electron (DrEl).  Although till now nobody knows what the
elementary particle (ElmPrt) means, there exist a possibility for a obvious
consideration of the unusual behaviour of all the relativistic quantized
micro particle (QntMcrPrt) by means of our transparent surveyed PhsMdl of the
DrEl. The PhsMdl of the DrEl is offered in all my work in resent nineteen for
bring of light to physical cause of the uncommon relativistic quantum
behaviour of the DrEl and give the thru physical interpretation and sense of
its dynamical parameters.  It turns up that all the relativistic dynamical
properties of the DrEl are results of the participate of its fine spread
(FnSpr) elementary electric charge (ElmElcChrg) in the Schrodinger's
self-consistent fermion strong correlated vortical harmonic oscillations,
then all the quantized dynamical properties of the SchEl are results of the
participate of its well spread (WllSpr) ElmElcChrg in the nonrelativistic
Furthian quantized stochastic boson circular harmonic oscillations. It is
used as for a obvious teaching the occurred physical micro processes within
the investigated phenomena, so for doing them equal with the capacity of its
mathematical correct description by the mathematical apparatus of the both
the quantum mechanics (QntMch): the nonrelativistic (NrlQntMch) and
relativistic (RltQntMch).

 The object of this lecture is to discuss, explain and bring to light on
the physical interpretation of the nonrelativistic quantum behaviour of the
Schrodinger's electron (SchEl) and of the relativistic quantum behaviour of
the dynamical parameters of the Dirac's electron (DrEl). The PhsMdl of the
DrEl is proposed by me twenty years ago. This PhsMdl can equally explain as
the physical causes for its unusual classical stochastic and so the quantum
dualistic wave-corpuscular behaviour of SchEl. One gives a new cleared
picturesque physical interpretation with mother wit of the physical scene of
the relativistic dynamical parameters of DrEl. In our transparent surveyed
PhsMdl of the DrEl one will be regarded as some point like (PntLk) ElmElcChrg,
taking simultaneously part in the following four different motions: A) The
isotropic three-dimensional relativistic quantized (IstThrDmnRltQnt) Einstein
stochastic boson harmonic shudders (EinStchBsnHrmShds) as a result of
momentum recoils (impulse kicks), forcing the charged QntMcrPrt at its
continuous stochastical emissions and absorptions of own high energy (HghEnr)
virtual photons (VrtPhtns) by its PntLk ElmElcChrg). This jerky motion
display almost Brownian classical stochastic behaviour (BrnClsStchBhv) during
a small time interval $\tau_1$, much less then the period $T$ of the
IstThrDmnRltQnt EinStchBsnHrmShds and more larger then the time interval $t$
of the stochastically emission or absorption of the Hgh-Enr VrtPhtn by its
PntLk ElmElcChrg. In a consequence of such jerks along the IstThrDmnRltQnt
EinStchBznHrmShds "trajectory" the DrEl's PntLk ElmElbChrg takes form of the
fine spread (FnSpr) ElmElcChrg. B) The IstThrDmnRltQnt Schrodinger fermion
vortical harmonic oscillation motion (SchFrmVrtHrmOscMtn) of the DrEl's FnSpr
ElmElcChrg. In a consequence of such jerks along the EinStchHrmOscMtn
"trajectory" the "trajectory" of the DrEl's FnSpr ElmElcChrg, participating
in the IstThrDmnRltQnt SchFrmVrtHrmOscMtn takes a strongly broken shape. Only
after the correspondent averaging over the "trajectory" of the IstThrDmnRltQnt
EinStchBsnHrmShdMtn we may obtain the cylindrically spread "trajectory" of
the IstThrDmnRltQnt SchFrmVrcHrmOscs' one, having got the form of the crooked
figure of an eight. Only such a motion along a spread uncommon "trajectory"
of the DrEl's FnSpr ElmElcChrg could through a new light over the SchEl's
WllSpr ElmElcChrg's space distribution and over the spherical symmetry of
the SchEl's WllSpr ElmElcChrg. This self-consistent strongly correlated
IstThrDmnRltQnt SchFrmVrtHrmOscs' motion may be described correctly by
means of the four components of its total wave function (TtlWvFnc) $\Psi$
and four Dirac's matrices ; $\alpha_j\,(\gamma_j)$ and $\beta\,(\gamma_o)$.

\section{Description of the relativistic quantized behaviour of the DrEl.}

In this way we may do as well as make a possibility for making clear the
spinor character of such a movement and all its consequences as the proper
mechanical momentum (MchMmn)(spin) and the rest self-energy, fermion symmetry
and fermion statistics. Only in a result of the participating the FnSpr
ElmElcChrg in the IstThrDmnRltQnt SchFrmVrtHrmOscs' motion all the components
of the resultant self-consistent (RslSlfCns) ElcInt of own QntElcMgnFld may
be exactly compensated and have zero values and all the components of the
RslSlfCns MgnInt of own QntElcMgnFld would be doubled in a comparison with
the corresponding RslSlfCns values of the MgnInt of own ClsElcMgnFld of the
NtnClsMcrPrt with same WllSpr ElmElcChrg, fulfilling Einstein's relativistic
stochastic harmonic oscillations motion in conformity with the laws of the
Einstein's relativistic classical mechanics (RltClsMch). It is turn out that
in a result of own useful participation of the DrEl's FnSpr ElmElcChrg the
self-energy at a rest of its electromagnetic self-action (ElcMgnSlfAct)
between its continuously moving potential and vector-potential to be minimized
in the with own corresponding self-consistent way. It turns up that the
RslSlfCns values of the ElcInt and the MgnInt of own QntElcMgnFld of the DrEl
are generated by its FnSpr ElmElcChrg, editing incessantly Hgh-Enr StchVrtPhts
at different moments of the recent half period of time in its corresponding
ivarious spatial positions and absorbed it in the form of the ElcMgnSlfAct in
the point of the DrEl's FnSpr ElmElcChrg's instantaneous positions; C) The
isotropic three-dimensional nonrelativistic quantized (IstThrDmnNrlQnt)
Furthian stochastic boson vortical harmonic oscillations (FrthStchBsnVrtHrmOscs)
of the SchEl as a result of the permanent electric interaction (ElcIntAct) of
its WllSpr ElmElcChrg with the ElcItn of the resultant QntElcMgnFld of the
low energy (LwEnr) StchVrtPhtns, stochastically generated by dint of the
fluctuating vacuum (FlcVcm) in the form of exchanging StchVrtPhtns between
FlcVcm and its FnSpr ElmElcChrg. This Furthian quantized stochastic uncommon
behaviour of the SchEl with own participation in the random trembling motion
(RndTrmMtn) is very similar to Brownian classical stochastic behaviour of the
BrnClsMcrPrt with own participation in the RndTrmMtn. But in principle the
exact description of the resultant behaviour of the SchEl owing of its
participation in the IstThrDmnNrlQnt FrthStchBsnCrcHrmOsc motions could be
done only by means of the NrlQntMch's and nonrelativistic ClsElcDnm's laws.
D) The classical motion of the Lorentz's electron (LrEl) around an well
contoured smooth thin trajectory which is realized in a consequence of some
classical interaction (ClsIntAct) of its overspread (OvrSpr) ElmElcChrg, bare
mass or magnetic dipole moment (MgnDplMnt) with the intensity of some external
classical fields (ClsFlds) as in the Newton nonrelativistic classical mechanics
(NrlClsMch) and in the nonrelativistic classical electrodynamics (ClsElcDnm).

 In Order to understand the physical cause for the origin of the relativistic
characteristics of the DrEl and some special feature of its quantum behaviour
we have to investigate the participate of its FnSpr ElmElcChrg in the
IstThrDmnRltQnt Schrodinger fermion vortical harmonic oscillation motion
(SchFrmVrtHrmOscMtn), describing the inner structure of the SchEl. Therefore
we shall try in following to describe the IstThrDmnRltQnt SchFrmVrtHrmOscMtn
of the DrEl's FnSpr ElmElcChrg by means of the well-known mathematical
apparatus of the RltQntMch and to give a green light of a new physical
interpretation by virtue of the known language and conceptions of the
NrlClsMch. We begin our new description by writing the well-known linear
partial differential wave equation in partial derivative (LnrPrtDfrWvEqtPrtDrv)
of Dirac, describing the RltQntMch behaviour of the DrEl. As it is well-known
there exist different representations of the LnrPrtDfrWvEqtPrtDrv of Dirac
within the RltQntMch. For example we begin with the presentation  of its
symmetrical representation :
\begin{equation}\label{a1}
\,i\,\hbar\,\frac{\partial\Psi}{\partial t}\,=\,H_d\,\Psi\,=
\,C\,(\,\alpha_j\,{\hat p}_j\,)\,\Psi\,+\,m\,C^2\,\beta\,\Psi ;
\end{equation}

where matrices $ \alpha_j $ and $ \beta $ have the well-known form :
\begin{equation}\label{a2}
\,\alpha_j\,=\,\left|\,
\begin{array}{cc}
\tilde{0} & \tilde{\sigma}_j \\
\tilde{\sigma}_j & \tilde{0}
\end{array}\,\right| \quad {\rm and} \quad \beta\,=\,\left|\,
\begin{array}{cc}
\tilde{1} & \tilde{0} \\
\tilde{0} & -\tilde{1}
\end{array}\,\right|\,;
\end{equation}

 The total wave function (TtlWvFnc) $\Psi $ of the DrEl within the NrlQntMch,
satisfying the LnrPrtDfrWvEqtPrtDrv of Dirac (\ref{a1}), has four components
$(\,{\psi}_1,\,{\psi}_2,\,{\psi}_3,\,{\psi}_4\,)$. Dirac had secured Lorentz'
invariation of his LnrPrtDfrWvEqtPrtDrv by dint of the introduction of four
matrices $(\,{\alpha}_1,\,{\alpha}_2,\,{\alpha}_3,\,\beta\,)$.

 There exist also the standard representation :
\begin{equation}\label{b1}
\,i\,\hbar\,\frac{\partial\tilde{\Psi}}{\partial t}\,=
\,{\tilde H}_d\,\tilde{\Psi}\,=
\,C\,(\,\gamma_j\,{\hat p}_j\,)\,\tilde{\Psi}\,+
\,m\,C^2\,\gamma_o\,\tilde {\Psi} \quad;
\end{equation}

where matrices $\gamma_j$ and $\gamma_o$ have the well-known form :
\begin{equation}\label{b2}
\gamma_j\,= \left|\,
\begin{array}{cc}
\tilde{\sigma}_j & \tilde{0} \\
\tilde{0} & -\tilde{\sigma}_j
\end{array}
\,\right|\quad {\rm and} \quad \gamma_o\,=\,\left|\,
\begin{array}{cc}
\tilde{0} & \tilde{1} \\
\tilde{1} & \tilde{0}
\end{array} \,\right|\,
\end{equation}

 In order to understand the physical meaning of the both TtlWvFnc $\Psi$ and
$\tilde{\Psi}$ we must rewrite the both LnrPrtDfrWvEqnsPrtDvt of DrEl
(\ref{a1}) and (\ref{b1}) in their component forms. In such a way we may
write the following system of the motion equations :
\begin{equation}\label{ab1}
\,i\,\hbar\,\frac{\partial\varphi}{\partial t}\,=
\,C\,(\,\sigma_j\,\hat{P}_j\,)\,\chi\,+\,m\,C^2\,\varphi\,;
\,i\,\hbar\,\frac{\partial\chi}{\partial t}\,=
\,C\,(\,\sigma_j\,\hat{P}_j\,)\,\varphi\,-\,m\,C^2\,\chi\,;
\end{equation}

\begin{equation}\label{ab2}
\,i\,\hbar\,\frac{\partial\eta}{\partial t}\,=
\,C\,(\,\sigma_j\,\hat{P}_j\,)\,\eta\,+\,m\,C^2\,\lambda\,;
\,i\,\hbar\,\frac{\partial\lambda}{\partial t}\,=
\,-\,C\,(\,\sigma_j\,\hat{P}_j\,)\,\varphi\,+\,m\,C^2\,\eta\,;
\end{equation}

 The eqs.(\ref{ab1}) and (\ref{ab2}) have been written by virtue of the
following useful designations of the component sets of both the TtlWvFcs
$\Psi$ and $\tilde{\Psi}$ :
\begin{equation}\label{ab3}
\,\varphi\,=\,\left|\,
\begin{array}{c}
\psi_1 \\
\psi_2
\end{array}\,\right|\,; \quad {\rm and} \quad
\,\chi\,=\,\left|\,
\begin{array}{c}
\psi_3 \\
\psi_4
\end{array}\,\right|\,; \quad {\rm or} \quad
\,\eta\,=\,\left|\,
\begin{array}{c}
\tilde{\psi}_1 \\
\tilde{\psi}_2
\end{array}\,\right|\,; \quad {\rm and} \quad
\,\lambda\,=\,\left|\,
\begin{array}{c}
\tilde{\psi}_3 \\
\tilde{\psi}_4
\end{array}\,\right|\,;
\end{equation}

 As it is following from (\ref{ab1}) the total energy of the DrEl at its
forward motion $i\,\hbar\,\frac{\partial\varphi}{\partial t}$ is equal of
the sum of its kinetic energy $C\,(\tilde{\sigma}_j\,\tilde{P}_j)$ in the
state $,\chi$ and potential energy $m.C^2$ of its ElcMgnSlfAct of its FnSpr
ElmElcFld with a corresponding resultant own QntElcMgnFld, created by the
StchVrtPhtns, radiated by its FnSpr ElmElcChrg in the state $\varphi$.
Just in such a way about the total energy of the DrEl $i\,\hbar\,
\frac{\partial\chi}{\partial t}$ is equal of the sum of its kinetic energy
$C\,(\tilde{\sigma}_j\,\tilde{P}_j)$ in the state $\varphi$ and potential
energy $m.C^2$ of its ElcMgnSlfAct of its FnSpr ElmElcFld with a
corresponding resultant own QntElcMgnFld, created by the StchVrtPhtns,
radiated by its FnSpr ElmElcChrg in the state $\chi$. That is because we may
flatly assert that the components $\varphi$ of its TtlWvFnc $\Psi$ in the
symmetrical presentation describe the forward motion of the DrEl and the
components $\chi$ of the same TtlWvFnc describe the backward motion of the
DrEl. Besides that if the both odd components:  $\Psi_1$ and $\Psi_3$
describe the DrEl's spinning in a left, then both even components : $\Psi_2$
and $\Psi_4$ describe the DrEl's spinning in a right. Therefore these two
group WvFnc have opportunity sign before the angle $\varphi$ are transforming
separately.

 The existence relations between both TtlWvFnc $\Psi$ and $\tilde{\Psi}$ give us a clear physical explanation and correct mathematical
description of all components of the DrEl's TtlWvFnc in the both
representation. From the LnrPrtDfrWvEqtPrtDrv of Dirac we may see that
matrixes work within its as switches, making possible the correct
mathematical description of the IstThrDmnRltQnt SchFrmVrtHrmOscMtn of the
DrEl's FnSpr ElmElcChrg by dint of the four components of its TtlWvFnc.

 In a due course it is easily to show further that the proper mechanical
moment (PrpMchMmn)(spin) of the DrEl can really be created as a result of the
participate of its FnSpr ElmElcChrg in the IstThrDmnRltQnt SchFrmVrtHrmOscMtn.
In order to obtain this in a naturally way we have to rewrite the
four one-component LnrPrtDfrWvEqtPrtDrv of Dirac (\ref{ab1}) in more obvious
form :
\begin{eqnarray}
\,i\,\hbar\,\frac{\partial {\psi}_1}{\partial t}\,+\,i\,\hbar\,C\,\times\,
\left\{\,\frac{\partial {\psi}_4}{\partial x}\,-\,i\,\frac{\partial {\psi}_4}
{\partial y}\,+\,\frac{\partial {\psi}_3}{\partial z}\,\right\}\,=
\,m\,C^2\,{\psi}_1\,;  \nonumber \\
\,i\,\hbar\,\frac{\partial {\psi}_2}{\partial t}\,+\,i\,\hbar\,C\,\times\,
\left\{\,\frac{\partial {\psi}_3}{\partial x}\,+\,i\,\frac{\partial {\psi}_3}
{\partial y}\,-\,\frac{\partial {\psi}_4}{\partial z}\,\right\}\,=
\,m\,C^2\,{\psi}_2\,;  \nonumber \\
\,i\,\hbar\,\frac{\partial {\psi}_3}{\partial t}\,+\,i\,\hbar\,C\,\times\,
\left\{\,\frac{\partial {\psi}_2}{\partial x}\,-\,i\,\frac{\partial {\psi}_2}
{\partial y}\,+\,\frac{\partial {\psi}_1}{\partial z}\,\right\}\,=
\,-\,m\,C^2\,{\psi}_3\,;  \nonumber \\
\,i\,\hbar\,\frac{\partial {\psi}_4}{\partial t}\,+\,i\,\hbar\,C\,\times\,
\left\{\,\frac{\partial {\psi}_1}{\partial x}\,+\,i\,\frac{\partial {\psi}_1}
{\partial y}\,-\,\frac{\partial {\psi}_2}{\partial z}\,\right\}\,=
\,-\,m\,C^2\,{\psi}_4\,;  \label{aa1}
\end{eqnarray}

 Then the system of four one-component LnrPrtDfrWvEqtPrtDrv of Dirac
(\ref{ab1}) may be rewritten by means of the substitution : $ x\,=\,
\rho \cos \phi $ and $ y\,=\,\rho \sin \phi\,$; from Decart's coordinates
in more obvious form of cylindrical coordinates :
\begin{eqnarray}
\,i\,\hbar\,\frac{\partial {\psi}_1}{\partial t}\,+\,i\,\hbar\,C\,
\left[\,\exp{-i\phi}\,\left\{\,\frac{\partial {\psi}_4}{\partial\rho}\,
-\,\frac{i}{\rho}\,\frac{\partial {\psi}_4}{\partial\phi}\,\right\}\,+
\,\frac{\partial {\psi}_3}{\partial z}\,\right]\,=\,m\,C^2\,{\psi}_1\,;
\nonumber \\
\,i\,\hbar\,\frac{\partial{\psi}_2}{\partial t}\,+\,i\,\hbar\,C\,
\left[\,\exp{+i\phi}\,\left\{\,\frac{\partial {\psi}_3}{\partial\rho}\,
+\,\frac{i} {\rho}\,\frac{\partial {\psi}_3}{\partial\phi}\,\right\}\,+
\,\frac{\partial {\psi}_4}{\partial z}\,\right]\,=\,m\,C^2\,{\psi}_2\,;
 \nonumber \\
\,i\,\hbar\,\frac{\partial {\psi}_3}{\partial t}\,+\,i\,\hbar\,C\,
\left[\,\exp{-i\phi}\,\left\{\,\frac{\partial {\psi}_2}{\partial\rho}\,
-\,\frac{i}{\rho}\,\frac{\partial {\psi}_2}{\partial \phi}\,\right\}\,+
\,\frac{\partial {\psi}_1}{\partial z}\,\right]\,=\,-\,m\,C^2\,{\psi}_3\,;
 \nonumber \\
\,i\,\hbar\,\frac{\partial {\psi}_4}{\partial t}\,+\,i\,\hbar\,C\,
\left[\,\exp{+i\phi}\,\left\{\,\frac{\partial {\psi}_1}{\partial\rho}\,
+\,\frac{i}{\rho}\,\frac{\partial {\psi}_1}{\partial\phi}\,\right\}\,-
\,\frac{\partial {\psi}_2}{\partial z}\,\right]\,=\,-\,m\,C^2\,{\psi}_4\,;
\label{aa2}
\end{eqnarray}

 As it is easily to seen if the first pair of TtlWvFnc's components
${\psi}_1$ and ${\psi}_3$ have equal phase factors ${\phi}_1$, than the
second the pair of TtlWvFnc's components ${\psi}_2$ and ${\psi}_3$ have also
equal phase factors $-{\phi}_1$. As it follows from eq.(\ref{aa2}) the
difference between two phase factors is equal of $\phi$. Therefore we may
suppose by means of a symmetrical consideration that four components
(OrbWvFnc) of the DrEl's TtlWvFnc $\Psi$ have the following presentations :
\begin{eqnarray}
{\psi}_1\,(\rho,\phi,z)\,=\,{\bar \psi}_1\,(\rho,\phi,z)\,\exp{-i(\phi/2)}\,;
\quad
{\psi}_3\,(\rho,\phi,z)\,=\,{\bar \psi}_3\,(\rho,\phi,z)\,\exp{-i(\phi/2)}\,;
\label{aa3} \\
{\psi}_2\,(\rho,\phi,z)\,=\,{\bar \psi}_2\,(\rho,\phi,z)\,\exp{+i(\phi/2)}\,;
\quad
{\psi}_4\,(\rho,\phi,z)\,=\,{\bar \psi}_4\,(\rho,\phi,z)\,\exp{+i(\phi/2)}\,;
\label{aa4}
\end{eqnarray}

 If we take into account that the participate of the well spread (WllSpr)
ElmElcChrg of the SchEl in the IstThrDmnRltQnt FrthStchBsnCrcHrmOscMtn,
securing its quantum behavior secures an additional dispersion
$({\delta {\phi}}/2)$, then the TtlWvFnc $\Psi$ of the DrEl can be rewritten
in two following representations :
\begin{eqnarray}
\Psi_{l+1/2}\,(\rho,\phi,z)\,=\,\frac{\phi_o}{2}\,\left|\,
\begin{array}{c}
\psi_{1l}\,(\rho,z)\,\exp{il\phi} \\
\psi_{2l}\,(\rho,z)\,\exp{i(l+1)\phi} \\
\psi_{3l}\,(\rho,z)\,\exp{il\phi} \\
\psi_{4l}\,(\rho,z)\,\exp{i(l+1)\phi} \\
\end{array}\,\right|\,\label{aa5}
\end{eqnarray}

and
\begin{eqnarray}
\Psi_{l-1/2}\,(\rho,\phi,z)\,=\,\frac{\phi_o}{2}\,\left|\,
\begin{array}{c}
\psi_{1l}\,(\rho,z)\,\exp{i(l-1)\phi} \\
\psi_{2l}\,(\rho,z)\,\exp{il\phi} \\
\psi_{3l}\,(\rho,z)\,\exp{i(l-1)\phi} \\
\psi_{4l}\,(\rho,z)\,\exp{il\phi} \\
\end{array}\,\right|\,\label{aa6}
\end{eqnarray}

 In the meanwhile it is easily to verify by virtue of the operator
\begin{equation}\label{aa7}
{\hat J}_z\,=\,\left\{\,-\,i\,\hbar\,\frac{\partial }{\partial \phi}\,+
\,\frac{\hbar}{2}\,\sigma_z\,\right\}
\end{equation}

that if the TtiWvFnc (\ref{aa5}) describes the behaviour of the free DrEl,
having the TtiMchMmn's value $J_z\,=\,\hbar\,(l+1/2)$, than the TtiWvFnc
(\ref{aa6}) describes the behaviour of the free DrEl, having the TtiMchMmn's
value $J_z\,=\,\hbar\,(l-1/2)$. Indeed, if
\begin{eqnarray}
{\hat J}_z\,\Psi_{l+1/2}\,(\rho,\phi,z)\,=\,\frac{\phi_o}{2}\,\left|\,
\begin{array}{c}
\hbar\,(l+1/2)\,\psi_{1l}\,(\rho,z)\,\exp{il\phi} \\
\hbar\,(l+1-1/2)\,\psi_{2l}\,(\rho,z)\,\exp{i(l+1)\phi} \\
\hbar\,(l+1/2)\,\psi_{3l}\,(\rho,z)\,\exp{il\phi} \\
\hbar\,(l+1-1/2)\,\psi_{4l}\,(\rho,z)\,\exp{i(l+1)\phi} \\
\end{array}\,\right|\,=\,\hbar\,(l+1/2)\,\Psi_{l+1/2}\,(\rho,\phi,z)
\label{aa8}
\end{eqnarray}

and
\begin{eqnarray}
{\hat J}_z\,\Psi_{l-1/2}\,(\rho,\phi,z)\,=\,\frac{\phi_o}{2}\,\left|\,
\begin{array}{c}
\hbar\,(l-1+1/2)\,\psi_{1l}\,(\rho,z)\,\exp{i(l-1)\phi} \\
\hbar\,(l-1/2)\,\psi_{2l}\,(\rho,z)\,\exp{il\phi} \\
\hbar\,(l-1+1/2)\,\psi_{3l}\,(\rho,z)\,\exp{i(l-1)\phi} \\
\hbar\,(l-1/2)\,\psi_{4l}\,(\rho,z)\,\exp{il\phi} \\
\end{array}\,\right|\,=\,\hbar\,(l-1/2)\,\Psi_{l-1/2}\,(\rho,\phi,z)
\label{aa9}
\end{eqnarray}

 The presented upper investigation shows us that the FnSpr ElmElcChrg of the
DrEl really participates in the IstThrDmnRltQnt SchFrmVrtHrmOscMtn and the
WllSpr ElmElcChrg of the SchEl really participates in the IstThrDmnRltQnt
FrthStchBsnCrcHrmOscMtn. In what following we wish to show that the
participating the FnSpr ElmElcChrg in its self-consistent IstThrDmnRltQnt
SchFrmVrtHrmOscMtn (Zitterbevegung) is very effective at creation of its own
resultant self-consistent quantized electromagnetic field (QntElcMgnFld) by
means of stochastic emission of the VrtPhtns.

 In the first Breit and afterwards Fock had observed that the instantaneous
velocity operator $\frac{\partial {\hat r}_j}{\partial t}$ assume, that the
operator of the instantaneous velocity of a free QntMcrPrt have very paradoxical
form in the RltQntMch. Indeed, it is well-known from the RltQntMch that the
analytical operator form of the instantaneous velocity value of the free DrEl
may be obtained by virtue of the Heisenberg commutation relations between the
operators of its radius-vector $\hat{r}_j$ and Dirac's hamiltonian $H_d$. In
such the way we can obtain:
\begin{equation}\label{d}
\,\hat{V}_j\,=\,\frac{d\,{\hat r}_j}{d\,t}\,=
\,-\,\frac{i}{\hbar}\,(\,{\bar r}_j\,{\bar
H}_d\,-\,{\bar H}_d\,{\bar r}_j\,)\,= \,C\,\alpha_j ;
\end{equation}

 Seventy years ago Schrodinger had investigated the physical meaning of the
operators in the RltQntMch, describing the relativistic quantized behaviour of
the DrEl in the old Dirac picture, making use of its linear Hamiltonian,
LnrPrtDfrWvEqtPrtDrv of Dr and four component TtlWvFnc $\Psi(r,t)$. First of
all he had obtained the motion equation for Dirac's matrices :
\begin{equation}\label{e}
\,i\,\hbar\frac{\partial {\alpha_j}}{\partial t}\,=
\,(\,\alpha_j\,H_d\,-\,H_d\,\alpha_j\,)\,=
\,2\,(\,\alpha_j\,H_d\,-\,C\,{\hat p}_j\,)\,=
\,2\,(\,C\,{\hat p}_j\,-\,\alpha_j\,H_d\,)\,;
\end{equation}

 After replacing in the equation (\ref{e}) the matrices $\alpha_j$ by the
matrices $\eta_j$ according to the following equation :
\begin{equation}\label{f}
\,\eta_j\,=\,\left(\,\alpha_j\,-\,\frac{C\hat{p}_j}{H_d}\,I_o\,\right)\quad;
\end{equation}

 Schrodinger had obtained the oscillation equation for $\eta$ matrices. In
such a way he had obtained the following solution of eq.(\ref{d}) for
$r(t)_j$ :
\begin{equation}\label{g}
\,{\hat r}_j\,=\,{\hat a}_j\,+\,\frac{t C^2{\hat p}_j}{H_d}\,I_o\,+
\,\frac{\,i\,C\,\hbar}{H_d}\left(\,\alpha_j\,-\,\frac{C{\hat p}_j}{H_d}
\,I_o\,\right)_{t=o}\,\exp{\{\frac{2itH_d}{\hbar}\}} ;
\end{equation}

 From eq.(\ref{g}) we can see that the second term describes the classical
motion of the free ClsMcrPrt and one increases in a linear way with the
current velocity when the time is increasing. The last term $\eta$ in the
eq.(\ref{g}) describes the self-consistent inner motion of the FnSpr
ElmElcChrg of the DrEl, which had called Zitterbewegung by Schrodinger.
As would be understand the participation of the SchEl's WllSpr ElmElcChrg
in the IstThrDmnNrlQnt FrthStchBsnCrcHrmOscsMtm is not described by its
OrbWvFnc and therefore there are no own part in the coordinate operator
(\ref{g}). The participation of the DrEl's FnSpr ElmElcChrg in the
IstThrDmnRltQnt SchFrmVrtHrmOscMtn, which is well described by the Dirac's
matrices $\alpha_j$ and the four components of the DrEl's total wave
function (TtlWvFnc), directs us to construct a new matrix' RltQntElcDnm in
an accordance with the Maxwell's nonrelativistic ClsElcDnm, where the
classical motion of some NtnClsPrt is described with the smooth thin line.
Moreover, as it will be obvious in the following investigation there exists
a possibility to understand the physical reasons of the rest self-energy
origin and the physical interpretation of the LnrPrtDfrWvEqt of Dirac.

 Indeed, we shall demonstrate in the what following that because of the
participation of the DrEl's FnSpr ElmElcChrg in its IstThrDmnRltQnt
SchFrmVrtHrmOsc motion, there exists a possibility to calculate also the
instantaneous RslSlfCns values of all the components of the ElcInt $E_j$
and of the MgnInt $H_j$ of the own resultant QntElcMgnFld in the point of
its instantaneous position by means of the Dirac's matrices, describing
this self-consistent motion. As it was shown above the RslQwn QntElcMgnFld
is begotten by Hgh-Enr StchVrtPhtns, emitted by its FnSpr ElmElcChrg in
different time moments from corresponding different space positions, being
occupied by its FnSpr ElmElcChrg during the last half-period.

 Indeed, for long time (about one centaur) ago, thence R.Schwarzschild had
written the electro-kinetic term $\rho(\varphi\,-\,v.A)$ into the Lagranjian
density, it is well-known from the Maxwell ClsElcDnm that when the LrEl is
found in the external ClsElcMgnFld its canonical impulse $p_j$ amounts in two
parts :
\begin{equation}\label{h}
\,P_j\,=\,p_j^k\,+\,p_j^p\,=\,m\,v_j\,-\,\frac{e}{C}\,A_j\,;
\,W\,=\,E\,-\,\frac{e}{C}\,\varphi\,;
\end{equation}

 The first part $p_j^k\,=\,m\,v_j$ is the well-known kinematic momentum and
the second part $p_j^p\,=-\frac{e}{C}\,A_j$ is the potential momentum, which
the DrEl with his FnSpr ElmElcChrg have acquired when it was brought in some
external ClsElcMgnFld. Therefore the uncommon behaviour of the SchEl in the
external QntElcMgnFld may be described as the behavior of the free one, only
replacing the canonical impulse operator $\,{\hat p}_j\,=
\,-\,i\,\hbar\,\nabla_j\,$ with its generalized impulse $P_j$ and
generalized energy $W$, described in the following form (\ref{h}):
\begin{equation}\label{i}
\,{\hat P}_j\,=\,{\hat p}_j\,-\,\frac{e}{C}\,{\hat A}_j\,;
\,{\hat W}\,=\,{\hat E}\,-\,\frac{e}{C}\,{\hat {\varphi}}\,;
\end{equation}

with canonical impulse operator ${\hat P}_j\,=\,-\,i\,\hbar\,\nabla_j\,$
and canonical energy operator ${\hat W}\,=\,i\,\hbar\,\frac{\,\partial }
{\partial t} $. For the sake of the intrinsic symmetry it is convenient
to make use of the generalized impulse of its zero component $P_o$ instead
of $\frac{\hat W}{C}$ and of the mechanical impulse of its zero component
$p_o$ instead of $\frac{\hat E}{C}$. However it is easy to verify that if
all the components of the canonical impulses commute between them-selves :
\begin{eqnarray}\label{j}
\,{\hat P}_j\,{\hat P}_l\,-\,{\hat P}_l\,{\hat P}_j\,=\,\delta_{jl}\,;
\quad and \quad
{\hat P}_j\,{\hat P}_o\,-\,{\hat P}_o\,{\hat P}_j\,=\,\delta_{jo} \, ;
\end{eqnarray}

then all the components of the kinetic impulse don't commute between oneself.
It is turned out that the commutations between the mechanical (kinematic)
impulse generate the values of the ElcInt $E_j$ and the MgnInt $H_j$ of the
external ClsElcMgnFld, as it is well-known from the NrlQntMch. Indeed,if we
rewrite the eqs.$~(\ref{i})$ in the following form :
\begin{equation}\label{k1}
\,-\,i\,\hbar\,\nabla_j\,=
\,{\hat p}_j \,-\,\frac{e}{C}\,{\hat A}_j\,=
\,m\,{\hat v}_j\,-\,\frac{e}{C}\,{\hat A}_j\,;
\end{equation}

\,and
\begin{equation}\label{k2}
\,i\,\frac{\hbar}{C}\,\frac{\partial }{\partial t}\,=
\,{\hat p}_o \,-\,\frac{e}{C}\,{\hat A}_o\,=
\,\frac{\hat E}{C}\,-\,\frac{e}{C}\,{\hat {\varphi}}\,;
\end{equation}

then the four commutations between the four mechanical (kinematic) impulse
components generate the six values of the ElcInt $E_j$ and of the MgnInt
$H_j$ of the external ClsElcMgnFld, as it is well-known from the NrlQntMch :
\begin{equation}\label{l1}
\,{\hat p}_j\,{\hat p}_l\,-\,{\hat p}_l\,{\hat p}_j\,=
\,m\,{\hat v}_j\,m\,{\hat v}_l\,-\,m\,{\hat v}_l\,m\,{\hat v}_j\,=
\,i\,\hbar\,\frac{e}{C}\,\varepsilon_{jlk}\,H_k\,;
\end{equation}

\,and
\begin{equation}\label{l2}
\,{\hat p}_j\,{\hat p}_o,-\,{\hat p}_o\,{\hat p}_j\,=
\,m\,{\hat v}_j\,m\,{hat v}_o\,-\,m\,{\hat v}_o\,m\,{\hat v}_j\,=
\,i\,\hbar\,\frac{e}{C}\,\,E_j\,;
\end{equation}

 The six components values of the ElcInt $E_j$ and the MgnInt $H_j$ of the
external QntElcMgnFld are determined for the space position $r_j$ of the
WllSpr ElmElcChrg of the SchEl (from the NrlClsMch's point of view) at the
time moment t.Indeed, there aren't any mistake in our physical interpretation.
Indeed, we have to remember that the IstThrDmnNrlQnt FrthStchBsnCrcHrmOscs'
motion is a result of the permanent interaction of the WllSpr ElmElcChrg
of the SchEl with the ElcInt $E_j$ of the StchVrtPhtns, generated in the
neighbour area of its localization by the FlcVcm. Hence as the
IstThrDmnNrlQnt FrthStchBsnCrcHrmOscs'motion of the QntMcrPrt has no
a thin and smooth classical trajectory, which may be determined as a
solution of the Newton equation, then we are forced to take advantage of
results of the Heisenberg commutations (\ref{l1}) and (\ref{l2}). Moreover
we have no right to take into account the screening influence of the FlcVcm
over the ElcInt at the description of the ElcMgmSlfInt of the FnSpr
ElmElcChrg and MgnDplMmn of the DrEl with the RslSlfCns values of the
ElcInt and the MgnInt of own resultant QntElcMgnFld as we already use it
at taking into account its influence at the description of the SchEl's
participation in the IstThrDmnNrlQnt FrthStchBsnCrcHrmOsc's motion.

 Consequently, the vector-potential $\check{A}_j$ and potential $\check{V}_j$,
which takes part within the LnrPrtDfrWvEqt of Dirac don't contain any
contribution of the existent StchVrtPhtns within the FlcVcm. However on
other hand it is known from the RltQntMch too that the operator of the
instantaneous velocity $V_j$ of the free FnSpr DrEl has the exceptional
form $C\alpha_j$ as it is seen from eq.(\ref{d}). Therefore if we want
to describe correctly the IstThrDmnRltQnt SchFrmVrtHrmOsc's motion of the
free DrEl, who is moving by its instantaneous velocity ${\hat V}_j \,=
\,C\,\alpha_j\,$ within own resulting QntElcMgnFld by virtue of the
mathematical apparatus of the RltQntMch, we must take use of the well-known
Heisenberg commutation relations (\ref{l1}) and (\ref{l2}).  Therefore we
may suppose that the DrEl's generalized moments in the RltQntMch could be
determined by Dirac's matrices as they ware done the velocity was exchange
by $C\alpha_j$ in the same eqs.(\ref{g}). Then we have obtain the following
presentation of the DrEl's mechanical momentum components in the motionless
coordinate system of reference relatively to the centre of the space
distribution of the SchEl's WllSpr ElmElcChrg :
\begin{equation}\label{m}
\,\check{p}_o\,=\,m\,C\,\hat{I}\,;
\quad {\rm and} \quad \check{p}_j\,=\,m\,C\,\alpha_j\,;
\end{equation}

where $\check{I}$ denotes the unitar matrix. Hence the HsnCmtRlt between the
kinetic momentum components (\ref{m}) of the DrEl in the RltQntMch, which are
analogous of the HsnCmtRlt for the SchEl to eqs.(\ref{l1}) and (\ref{l2}).
have to determine the RslSlfCns values of the ElcInt ${\check E}_j$ and the
MgnInt ${\check H}_j$ of own QntElcMgnFld. Therefore they may be written in
the following form :
\begin{equation}\label{n1}
\,{\check p}_j\,{\check
p}_l\,-\,{\check p}_l\,{\check p}_j\,=
\,m^2\,C^2\,(\,\alpha_j\,\alpha_l\,-\,\alpha_l\,\alpha_j\,)\,=
\,2\,i\,m^2\,C^2\,\varepsilon_{jlk}\,\sigma_k\,;
\end{equation}

and
\begin{equation}\label{n2}
\,\check{p}_j\,\check{p}_o\,-\,\check{p}_o\,\check{p}_j\,=
\,m^2\,C^2\,(\,\alpha_j\,\hat{I}\,-\,\hat{I}\,\alpha_j\,)\,=\,0\,;
\end{equation}

if the commutations (\ref{l1}) and (\ref{l2}) between the mechanical
(kinematic) impulse ${\hat p}_j,\,{\hat p}_l$ and ${\hat p}_o$ generate the
AvrVls ${\hat E}_j$ and ${\hat H}_j$ of the ElcInt and MgnInt of the external
QntElcMgnFld, then the commutations (\ref{n1}) and (\ref{n2}) between the
mechanical (kinematic) impulse ${\check p}_j,\,{\check p}_l\,$ and
${\check p}_o$ generate the PslSlfCnsVls ${\check E}_j$ and ${\check H}_j$
of the ElcInt and the MgnInt of the own resultant QntElcMgnFld, which may be
obtained by means of the comparison of the corresponding right parts of the
(\ref{l1}), (\ref{l2}) and (\ref{n1}), (\ref{n2}) and one may be described
by the following formulas :
\begin{equation}\label{o}
\,{\check E}_j\,=\,0\,; \quad {\rm and} \quad
{\check H}_j\,=\,2\,\frac{m^2\,C^3}{e\,\hbar}\sigma_j\,;
\end{equation}

 In such an easy way only owing to a supposition of the self-consistency of
IstThrDmnRltQnt SchFrmVrtHrmOsc's motion of the DrEl's FnSpr ElmElcChrg at
about the light velocity C, who minimizes in a self-consistent way the
self-energy at a rest of its electromagnetic self-acting (ElcMgnSlfAct)
between its continuously moving FnSpr ElmElcChrg and the corresponding
electric current with their corresponding potential and vector-potential.
The RslSlfCnsVls of own QntElcMgnFld of the DrEl is created by its FnSpr
ElmElcChrg, emitting incessantly high energy StchVrtPhtns at different
moments of the latest time from its various corresponding positions in a
space and absorbed in the form of the ElcMgnSlfAct by the DrEl's FnSpr
ElmElcChrg in its instantaneous positions. All the RslSlfCnsVls of the
ElcInt components $\check{E}_j$ of own QntElcMgnFld may be precisely
compensated in a result of the participation of the DrEl's FnSpr ElmEl
in same IstThrDmnRltQnt SchFrmVrtHrmOscsMtn. Only in a result of such the
IstThrDmnRltQnt SchFrmVrtHrmOsc's motion all the RslSlfCns values of the
MgnInt components $\check{H}_j$ of own QntElcMgnFld may obtain double
values in a comparison with the corresponding averaged values of the
MgnInt components $\hat{H}_j$ of the ClsElcMgnFld of some NtnClsPrt,
charged by the FnSpr ElmElcChrg, fulfilled the IstThrDmnRltQnt
BsnStchCrcHrmOsc's motion in a conformity with the laws of the Einstein
relativistic classical mechanics (RltClsMch).

 The Schrodinger's zitterbewegung is a self-consistent strong correlated
vortical motion, which minimizes the self-energy at a rest of the QntMcrPrt
and one secures the continuous stability of its Shrodinger's wave package
(SchWvPct) in time within the space. Therefore this inner self-consistent
quantized vortical motion of the QntMcrPrt's FnSpr ElmElcChrg corresponds
to its inner harmonical motion, introduced firstly by Louis de Broglie.

There exists some possibilities enough not only for rough calculations of
the averaged undivergent potential and vector-potential of the DrEl's FnSpr
ElmElcChrg and real values of its particular MgnDplMm $\mu_j$ and MchMmn
$s_j$, which are results of the participation of the DrEl's FnSpr ElmElcChrg
in its IstThrDmnRltQnt SchFrmVrtHrmOsc's motion. This natural conclusion
explains the physical cause of the doubled gyromagnetical ratio of the
DrEl's inner MgnDplMmn $\mu_j\,=\,\frac{-e\hbar}{2mC}\,\sigma_j$ to its
inner MchMmn (spin) $S_j\,=\,\frac{\hbar}{2}\,\sigma_j$ and abolishes the
necessity of useless renormalization of its ElcChrg and mass because of
the absence of any physical substantiations, as will be seen in further
researches. Hence the magnetic productivity of the DrEl's FnSpr ElmElcChrg
as a result of its participation in the IstThrDmnRltQnt SchFrmVrtHrmOsc's
motion exceeds in twice the magnetic productivity of the SchEl's WllSpr
ElmElcChrg as a result of its participation of IstThrDmnNrtQnt
FrthStchBsnHrmOsc's motion with same parameters.

 In the proposed PhsMdl of the DrEl its rest energy $\,m.C^2\,$ may be
considered as a natural consequence to the ElcMgnSlfAct between its
RslSlfCns values of the MgnInt ${\check H}_j\,=
\,2(\frac{m^2C^3}{e\hbar})\,\sigma_j$ of the own QntElcMgnFld in the point
of its instantaneous position and the RslSlfCns values of its MgnDplMmn
$\mu_j\,=\,(\frac{-e\hbar}{2mC})\sigma_j$ at same point :
\begin{equation}\label{p}
\,E_o\,=\,-\,\langle\,{\mu}_j\,\rangle\,\langle\,\check{H}_j\,\rangle\,=
\,\frac{e\hbar}{2mC}\,\langle\,\sigma_j\,\rangle
\,\frac{2m^2\,C^3}{e\hbar}\,\langle\,\sigma_j\,\rangle\,=
\,mC^2\,\langle\,\sigma_j\,\rangle\,\langle\,\sigma_j\,\rangle\,=\,m\,C^2\,;
\end{equation}

 Indeed, it is easy to verify that :
\begin{equation}\label{q}
\,\langle\sigma_j\rangle\,\langle\sigma_j\rangle\,=
\,\sin^2\theta\,cos^2\phi\,+\,\sin^2\theta\,sin^2\phi\,+
\,cos^2\theta\,=\,1\,;
\end{equation}

 As a result of the above investigation we can affirm that the participation
of the SchEl's WllSpr ElmElcChrg in IstThrDmnNrlQnt FrthStchCrcBsnHrmOscMtn
cause existence of its anomalous MgnDplMmn $\delta{\mu}_o$. Therefore if we
wish to obtain a ratio of the anomalous MgnDplMmn $\,\delta{\mu}_o\,$ of the
SchEl's WllSpr ElmElcChrg as a result of its participate in the
IstThrDmnNrlQnt FrthStchCrcBsnHrmOscMtn to its own MgnDplMmn $\,\mu_o\,$ as a
result of the participation of DrEl's FnSpr ElmElcChrg in the IstThrDmnNrlQnt
SchFrmHrmOscMtn, then we must know that it is equal to half ratio of their
kinetic energies.  Consequently as a result of the natural relations we can
obtain :
\begin{equation}\label{r}
\,\frac{\delta{\mu}_o}{\mu_o}\,=\,\frac{1}{2\pi}\,\frac{e^2}{C\hbar}
\,\frac{mC^2}{mC^2}\,;
\quad and \quad
\delta{\mu}_o\,=\,\frac{\mu_o}{2\pi}\,\frac{e^2}{C\hbar}\,;
\end{equation}

 In a consequence of the above used approach we may propose that the
RslSlfCns values of the four components of the vector-potential
${\check A}_j$ and potential ${\check {\phi}}\,$ of the DrEl's FnSpr
ElmElcChrg, participating in the IstThrDmnNrlQnt SchFrmVrtHrmOsc's motion,
one have the following analytic form :
\begin{equation}\label{s}
\,\langle\,\check{A}_j\,\rangle\,=\,\frac{m.C^2}{-\,e}
\,\langle\,\Psi^*\,|\,\beta\,\alpha_j\,|\,\Psi\,\rangle\,;
\quad {\rm and} \quad
\langle\,\check{\phi}\,\rangle\,=\,\frac{m.C^2}{-\,e}
\langle\,\Psi^*\,|\,\beta\,|\,\Psi\,\rangle\,;
\end{equation}

 There exists a physical cause and a mathematical possibility for obtaining
the RslSlfCns values $(\ref{o})$ of the ElcInt ${\check E}_j$ and the MgnInt
${\check H}_j$ from the four components in eqs.(\ref{s}) of their
vector-potential ${\check A}_j$ and potential ${\check {\phi}}$ by means of
the corresponding relations between analogous ones of that, expressing the
Maxwell's laws within the Maxwell's ClsElcDnm. For that purpose we have to
exchange the coordinate operators $\nabla_j$ and time operator
$\frac{1}{C}\,\frac{\partial }{C\partial t}$ in the Maxwell's ClsElcDnm's
scheme, where the ElcChrg McrPrt's motion is described by its coordinate
$r_j$, by the matrix operators $i\,\frac{mC}{\hbar}\,\alpha_j$ and
$\frac{mC}{\hbar}\,\hat {I}$, when the FnSpr ElmElcChrg's motion is described
by the matrices, in accordance with the relations (\ref{l1}), (\ref{l2}),
(\ref{n1}), (\ref{n2}), and (\ref{o}). Indeed, then instead of the Maxwell's
relations :
\begin{equation}\label{t1}
\,H_j\,=\,\left[\,\nabla\,\times\,A\,\right]_j\,=
\,\varepsilon_{jkl}\,\nabla_k\,A_l\,;
\end{equation}

and
\begin{equation}\label{t2}
\,E_j\,=\,-\frac{1}{C}\,\frac{\partial A_j}{\partial t}\,-\,\nabla_j\,\phi\,
\end{equation}

we must write the following unusual equations :
\begin{equation}\label{u1}
\,{\check H}_j\,=\,i\,\varepsilon_{jkl}\,\frac{mC}{\hbar}\,\alpha_j\,
\frac{mC^2}{-e}\,\alpha_l\,=\,2\frac{m^2C^3}{e\,\hbar}\,\sigma_j\,;
\end{equation}

and
\begin{equation}\label{u2}
\,{\check E}_j\,=\,i\,\frac{mC}{\hbar}\,\hat{I}\,\frac{mC^2}{-e}\,\alpha_j\,-
\,i\,\frac{mC}{\hbar}\,\alpha_j\,\frac{mC^2}{-e}\,\hat{I}\,=\,0\,;
\end{equation}

 There is a necessity to note here that the presence of the matrices in the
(\ref{o}) is connected with the physical reason that the moment RslSlfCns
values of the potential and vector-potential in the time $t_1$ are results
of the interference of the QntElcMgnFlds of all Hgh-Enr StchVftPhtns,
emitted in the time interval $(t_1\,-\,\frac{T}{2})\,\leq\,t\,\leq\,t_1$.
It is easy to be shown in Veile's symmetrical representation, where the
matrices $\gamma_j$ and $\gamma_o$ play the parts of the matrices $\alpha_j$
and $\beta$, participating in the Pauli-Dirac's representation.

 As a consequence of our felicitious supposition, using by me at building
of our PhsMdl, we may write the LnrDfrWvEqnPrtDrv of Dirac by means of the
expression (\ref{s}) for the RslSlfCns values of the own potential
${\check {\phi}}$ and vector-potential ${\check A}_j$ of the DrEl's FnSpr
ElmElcChrg and its corresponding current in the following form :
\begin{eqnarray}\label{v}
\,\langle\,\Psi^{+},|\,H_d\,|\,\Psi\,\rangle\,=
\,-\,e\,\langle\,\Psi^{+}\,|\,\Psi\,\rangle\,\frac{m.C^2}{-e}
\,\langle\,\Psi^{+}\,|\,\beta\,|\,\Psi\,\rangle\,+      \nonumber\\
\,e\,C\,\langle\,\Psi^{+}\,|\,\alpha_j\,|\,\Psi\,\rangle\,\frac{mC}{e}
\,\langle\,\Psi^{+}\,|\,\beta\,\alpha_j\,|\,\Psi\,\rangle\,+
\,C\,\langle\,\Psi^{+}\,|\,\alpha_j\,P_j\,|\,\Psi\,\rangle\,;
\end{eqnarray}

 Then if we take into account the existence of the following equations\,:
\begin{equation}\label{w}
\,\langle\,\Psi^{+}\,|\,\Psi\,\rangle\,=\,1\,; \qquad {\rm and} \qquad
\,\langle\,\Psi^{+}\,|\,\alpha_j\,|\,\Psi\,\rangle
\,\langle\,\Psi^{+}\,|\,\beta\,\alpha_j\,|\,\Psi\,\rangle\,=\,0\,;
\end{equation}

then the LnrPrtDfrWvEqn of Dr. The (\ref{v}) may be rewritten into the
following well-known form :
\begin{equation}\label{ba1}
\,\langle\,\Psi^{+}\,|\,H_d\,|\,\Psi\,\rangle\,=
\,C\,\langle\,\Psi^{+}\,|\,\alpha_j\,P_j\,|\,\Psi\,\rangle\,+
\,m\,C^2\,\langle\,\Psi^{+}\,|\,\beta\,|\,\Psi\,\rangle\,
\end{equation}

 Our investigation shows the role of each component of the TtlWvFnc $\Psi$
of the DrEl at the description of its behaviour and the role of the matrixes
as useful switches, making possible the description of the DrEl's motion by
means of its four-component TtlWvFnc. Because of such our interpretation of
the space-time dependent part of the SchEl's one-component OrbWvFnc $\Psi$ ,
which gives the description of its orbital motion only, one could be compared
with black and white TV, while the space-time dependent part of the DrEl's
four-components TtlWvFnc $\Psi$, which together with matrices give the
description of its full motion,one should be compared with color TV. In other
words, while one-component OrbWvFnc $\psi$ give us the colourless description
only of the SchEl's motion, while the four-components TtlWvFnc $\Psi$
together with matrices give diversiform of the coloured description of the
DrEl's motion.

 The existence of the well-known relativistic relation $E^2\,=\,P^2C^2\,+
\,m^2C^4$ between the energy, momentum and mass of a real elementary particle
(ElmPrt) also may be obtained by the Maxwell equations of the ClsElcDnm,
taking into consideration the MgnInt between the momentum of the charged
ClsMcrPrt and the vector-potential of its ClsElcMgnFld, participating in the
IstThrDmnNrlQnt StchBsnVrtHrmOscMtn

 In the long run I cherish hope that our consideration of the massive
leptons' behaviour from a new point of view of my felicitous PhsMdl,
physically obvious expatiating the physical cause for the origin of all
their properties, will be of great interest for all scientists.
Theoretical physicists will find the picturesque explanation of all their
properties and Philosophers will find the total employment of the dialectical
materialism for the investigations of the micro particle's world. My very
quality and profound interpretation of the SchEl's behaviour in the NrlQntMch
and of the DrEl's behaviour in the RltQntMch , as well as the excellent
quantity coincidence of the deduced values of all DrEl's parameters with
their corresponding experimentally determined values gives as the hope for
correctness of our beautiful, simple and preposterous PhsMdl and fine,
extraordinary ideas, which have been inserted at its construction.


\vspace{0.6cm}

\begin{thebibliography}{99}

\bibitem{JMRa} Rangelov J.M.,Reports ofJINR, R4-80-493; R4-80-494; (1980),
Dudna.

\bibitem{JMRb} Rangelov J.M., University Annual (Technical Physics), {\bf 22},
(2), 65, 87, (1985); {\bf 23}, (2), 43, 61, (1986); {\bf 24}, (2), 287, (1986);
{\bf 25}, (2), 89, 113, (1988).

\bibitem{JMRc} Rangelov J.M.,Comptens Rendus e l'Academie Bulgarien Sciences,
{\bf 39}, (12), 37, (1986).

\bibitem{LB} De Broglie L.,Comptens Rendus  {\bf 177},507, 548, 630, (1923);
Ann. de Physique  {\bf 3}, 22 (1925).

\bibitem{WH} Heisenberg W., Ztschr.f.Phys. {\bf 33}, 879, (1925) ; {\bf 38},
411, (1926); {\bf 43}, 172, (1927). Mathm.Annalen {\bf 95}, 694, (1926);

\bibitem{CGD} Darvin C.G., Proc.Roy.Soc.(L), {bf 115}, 1, (1927); {\bf 115},
1, (1927); {\bf 117}, 258, (1927); The New Conceptions of Matter, McMillan,
N.Y. (1931).

\bibitem{WPa} Pauli W., Ztschr.f.Phys., {\bf 31}, 765, (1925) ; {\bf 36}, 336,
(1926); {\bf 41}, 81, (1927); {\bf 43}, 601, (1927).

\bibitem{PAMa} Dirac P.A.M.,Proc.Cambr.Phil.Soc., {\bf 22}, 132, (1924);
 Proc.Roy.Soc.(L), {\bf A106}, 581, (1924);  {\bf A112}, 661, (1926);
{\bf A113}, 621, (1927).

\bibitem{PAMb} Dirac P.A.M.,Proc.Roy.Soc.(L),  {\bf A117}, 610; {\bf A118},
 127, 351, (1928);  {\bf68}, 527, (1931).

\bibitem{DBr} Breit D., Proc.Nat.Acad.Scien.USA {\bf 14}, 553, (1928);
{\bf 17}, 70, (1931).

\bibitem{VAF} Fock V.A., Ztscr.f.Physik {\bf 55}, 127, (1928); {\bf 68}, 527,
(1931)

\bibitem{ESchc} Schrodinger E., Sitzungber.Preuss.Akad.Wiss., {\bf K1}, 296,
418, (1930).

\bibitem{AE} Einstein A., Ann.d.Phys., {\bf 17}, 549, (1905); {\bf 19}, 371,
(1906);  {bf 33}, 1275, (1910);

\bibitem{ThW} Welton Th.,Phys.Review,  {\bf 74}, 1157, (1948)

\bibitem{JMRd} Rangelov J.M., Report Series of Symposium on the Foundations
of Modern Physics, 6/8, August, 1987, Joensuu,p.95-99, FTL, 131, Turqu,
Finland; Problems in Quantum Physics'2, Gdansk'89,  18-23, September, 1989,
Gdansk, p.461-487,  World Scientific, Singapur, (1990);

\bibitem{JMRe} Rangelov J.M., Abstracts Booklet of 29th Anual Conference of
the University of Peoples' Friendship, Moscow 17-31 May 1993  Physical ser.;
Abstracts Booklet of Symposium on the Foundations of Modern Physics,  13/16,
June,(1994),  Helsinki, Finland  60-62.

\bibitem{JMRf} Rangelov J.M., Abstract Booklet of B R U-2,  12-14,
September, (1994), Ismir, Turkey ; Balk.Phys.Soc.,  {\bf 2}, (2), 1974,
(1994).  Abstract Booklet of B R U -3,  2-5  September,(1997),
Cluj-Napoca,Romania.

\bibitem{NK} Kalitcin N.,(in Russian), JETPH  {\bf 25}, 407, (1953).

\bibitem{ASBT} Socolov A.A., Tumanov V.S.,(in Russian) JETPH,  {\bf 30},
 802, (1956).

\bibitem{AAS} Sokolov A.A.,Scientific reports of higher school,(in Russian)
 (1), 120, (1963) Moscow ;  Phylosophical problems of elementary particle
physics.(in Russian), Acad.of Scien.of UdSSR ,Moscow,  188,  (1963).

\end{thebibliography}
\end{document}